\documentclass[twocolumn,prl,amsmath,amssymb]{revtex4-1}
\usepackage[latin9]{inputenc}
\usepackage{amsmath}
\usepackage{amssymb}
\usepackage{graphicx}
\usepackage[unicode=true,pdfusetitle,
 bookmarks=true,bookmarksnumbered=false,bookmarksopen=false,
 breaklinks=false,pdfborder={0 0 1},backref=false,colorlinks=false]
 {hyperref}
\begin{document}
\title{Emergent criticality and universality class of spin and charge density
wave transitions of two-component lattice Bose gases in optical cavities
at finite temperature}
\author{Liang He}
\email{liang.he@scnu.edu.cn}

\affiliation{Guangdong Provincial Key Laboratory of Quantum Engineering and Quantum
Materials, SPTE, South China Normal University, Guangzhou 510006,
China}
\author{Su Yi}
\email{syi@itp.ac.cn}

\affiliation{CAS Key Laboratory of Theoretical Physics, Institute of Theoretical
Physics, Chinese Academy of Sciences, Beijing 100190, China}
\affiliation{School of Physical Sciences \& CAS Center for Excellence in Topological
Quantum Computation, University of Chinese Academy of Sciences, Beijing
100049, China}
\begin{abstract}
We investigate the finite temperature spin density wave (SDW) and
charge density wave (CDW) transition of two-component lattice spinor
Bose gases in optical lattices in the Mott-insulator limit. At the
temperature scale around half of the on-site interaction energy, we
find a new critical regime emerges and features, in particular, a
new bicritical line and two critical lines associated with the finite
temperature SDW-CDW, homogeneous-SDW, and homogeneous-CDW transition,
respectively. Direct calculation of the critical exponents for the
scaling behavior and investigating on the effective theory in this
critical regime show that they belong to the five-dimensional Ising
universality class, clearly manifesting the long-range character of
the system's interaction. Our prediction of the emergent criticality
can be readily observed by current experimental setups operated at
the intermediate temperature scale around half the on-site interaction
energy. 
\end{abstract}
\maketitle
Many-body systems with high microscopic symmetry can give rise to
the rich interplay between various macroscopic orders. Well-known
examples in the context of conventional condensed matter physics,
range from the multicritical points associated with various magnetic
orders found in antiferromagnets with weak lattice anisotropy \citep{Shapira_PR_1970_bicritical_point,Rohrer_PRL_1977_bi_and_tetracritical_point},
over the multiferroics where the magnetic order and the electric order
are coupled in a non-trivial way holding the potential in greatly
improving the energy efficiency of electronic devices \citep{Spaldin_Phys_Today_2010},
to the even more elusive scenario in high-temperature superconductors
where the charge density wave (CDW), spin density wave (SDW), and
superconducting order can compete, intertwine, or even form the ``vestigial
order'' in certain cases \citep{Fradkin_RMP_2015}. Noticing, however,
interactions in these conventional condensed matter systems are all
short-ranged, while in the context of ultracold atom physics, not
only short-range interacting, but also long-range interacting quantum
many-body systems can be realized and well-controlled in experiments.
This thus provides an ideal platform for investigating the rich interplay
between various macroscopic orders in the presence of long-range interactions. 

Among various long-range interacting ultracold atom systems, Bose
gases in optical cavities are unique in the sense that they assume
\emph{infinite-long range} (ILR) interactions. In particular, ultracold
Bose gases with internal spin degrees of freedom have recently been
realized in experiments \citep{Landini_PRL_2018,Morales_Nat_Mat_2018,Kroeze_PRL_2018,Morales_PRA_2019,Dogra_Science_2019,Davis_PRL_2019,Li_arXiv_2020},
featuring both an ILR spin-spin and an ILR density-density interaction
mediated by cavity photons. In the experiments of Ref.~\citep{Landini_PRL_2018,Dogra_Science_2019},
two macroscopic orders, namely the SDW and the CDW order, have been
observed and a first-order transition between these two ordered phases
have been identified by tuning the ratio between these two types of
ILR interactions. Noticing current experiments have mainly focused
on the low-temperature regime where the temperature is much smaller
than all other energy scales in the system, it is intriguing to expect
that at a comparable temperature scale, a new scenario for the interplay
between the two macroscopic orders could emerge. This thus gives rise
to the fundamental question of the existence and the universality
class of the possible new criticality that is absent in the low-temperature
regime, in particular, the multicritical behavior associated with
the interplay between the SDW and the CDW order.

\begin{figure}
\includegraphics[height=1.8in]{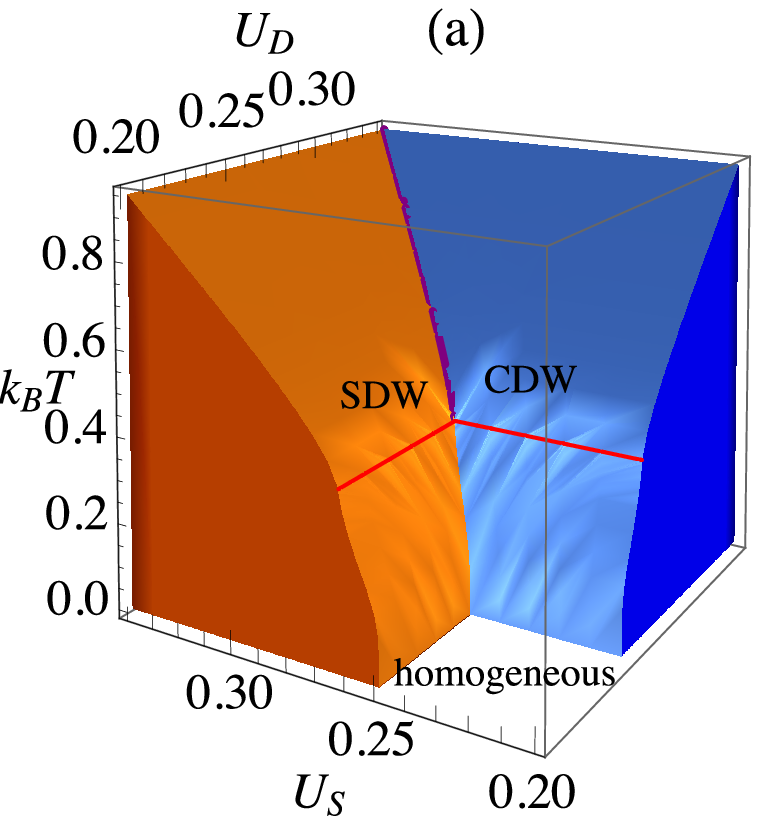}\includegraphics[height=1.75in]{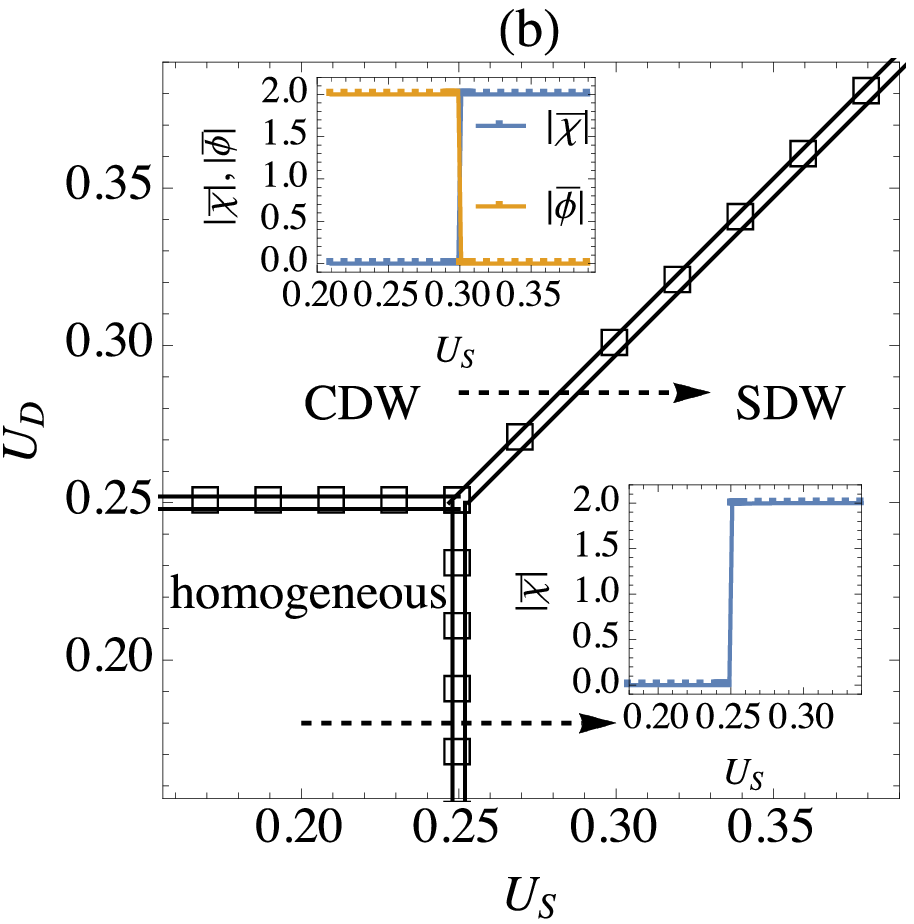}

\includegraphics[height=1.75in]{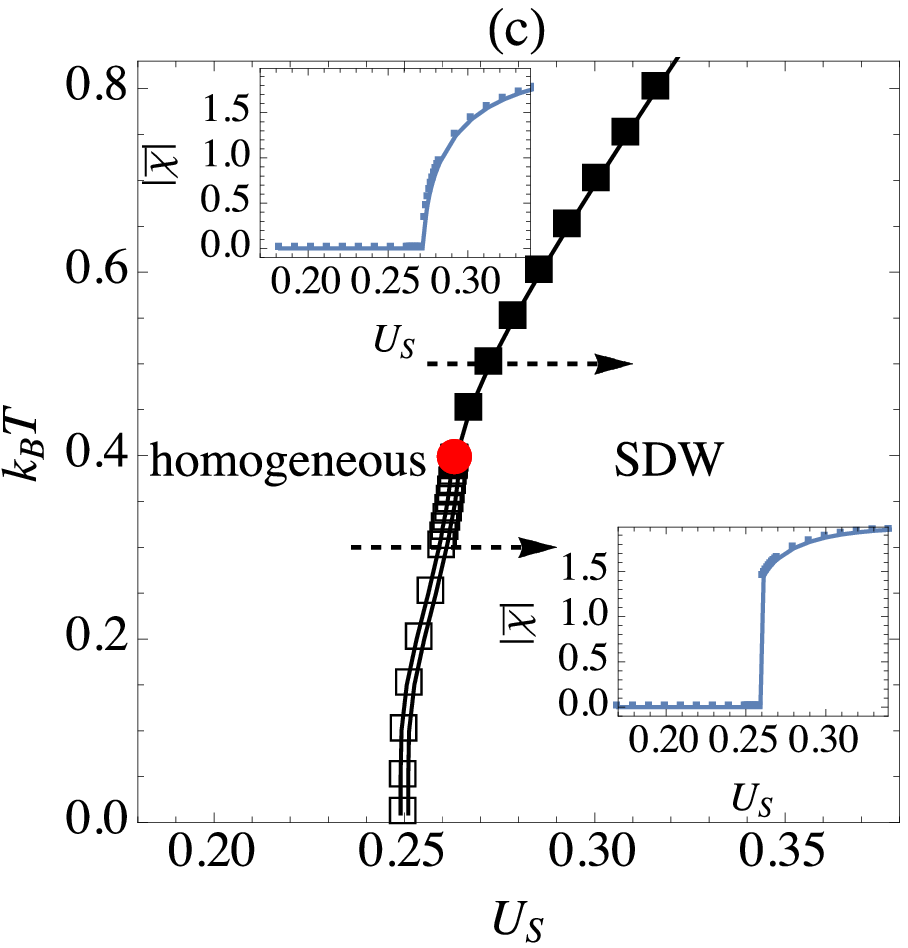}\includegraphics[height=1.75in]{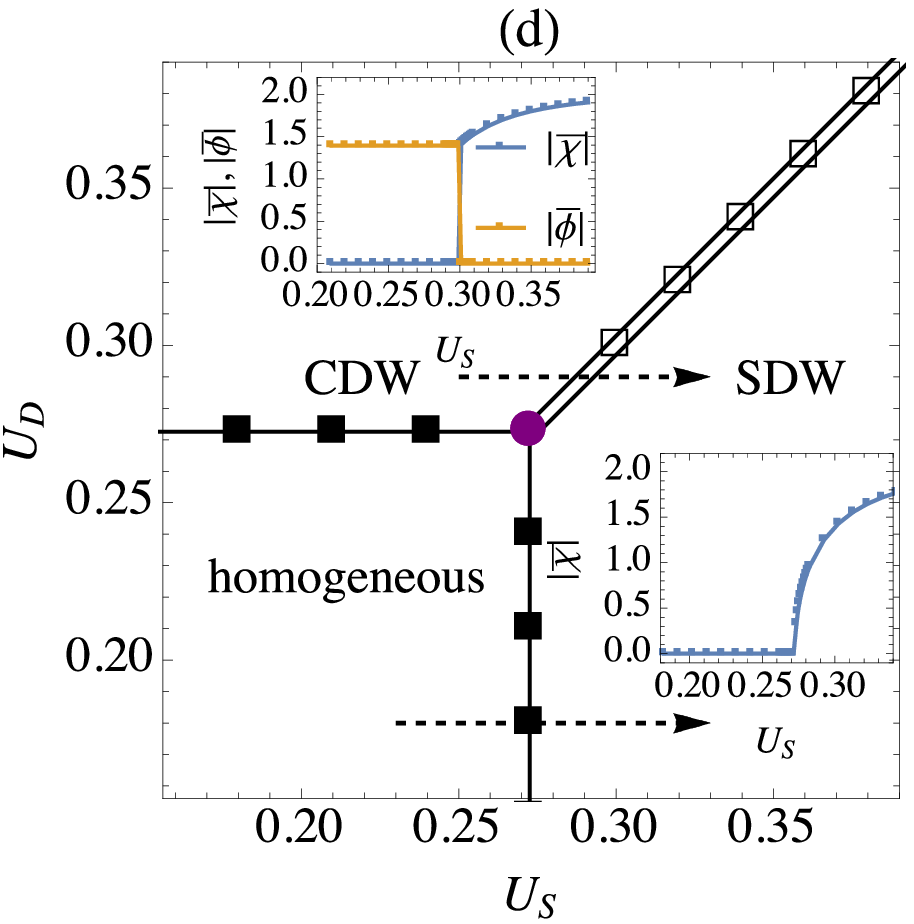}

\caption{(a) Finite temperature phase diagram of the system at the balanced
unit filling for both species of atoms, i.e., $\langle\sum_{i}\hat{n}_{i,\sigma}\rangle/L=1$.
$U_{+-}\ll U_{++}=U_{--}$ is assumed, and $U_{\sigma\sigma}$ is
set to be the energy unit. The SDW and CDW phase regions are highlighted
by orange and blue, respectively. Two critical lines (marked by the
thick red lines) emerge at the temperature $k_{B}T=0.395$. The left
(right) critical line separates the first-order homogeneous-SDW (CDW)
transition from the second-order homogeneous-SDW (CDW) transition.
Above these two critical lines, both the homogeneous-SDW and the homogeneous-CDW
transition are second-order transitions, and their transition boundaries
(second-order critical surfaces) meet at a bicritical line (marked
by the thick purple line). (b) Phase diagram at the fixed temperature
with $k_{B}T=0.1$. The transitions between each two of the homogeneous,
CDW and SDW phase are all first-order transitions whose boundaries
are marked by double solid curves and open squares. Insets: The order
parameter dependence on $U_{S}$ when the first-order CDW-SDW (homogeneous-SDW)
transition boundary is crossed. (c) Phase diagram at the fixed $U_{D}=0.2$.
At low temperatures, the homogeneous-SDW transition is a first-order
transition, whose boundary is marked by open squares and a double
solid curve. It becomes a second-order transition (marked by filled
squares and a solid curve) at and above a critical point (marked by
the red dot) whose temperature $T_{\mathrm{CP}}=0.395/k_{B}$. Insets:
The upper (lower) inset shows the SDW order parameter dependence on
$U_{S}$ with $k_{B}T=0.3$ ($k_{B}T=0.5$), clearly showing the transition
is a first-order (second-order) transition for $T<T_{\mathrm{CP}}$
($T>T_{\mathrm{CP}}$). (d) Phase diagram at the fixed temperature
with $k_{B}T=0.5$. The SDW-CDW transition still keeps as a first-order
transition, whose boundary is marked by a double solid curve and open
squares, while it terminates at a bicritical point (marked by the
purple dot), where the two second-order transition boundaries meet
(marked by solid curves and solid squares). Insets: The order parameter
dependence on $U_{S}$ when the first-order CDW-SDW (the second-order
homogeneous-SDW) transition boundary is crossed. See text for more
details.}
\label{Fig_1_Phase_diagram}
\end{figure}

\begin{figure}
\includegraphics[width=1.6in]{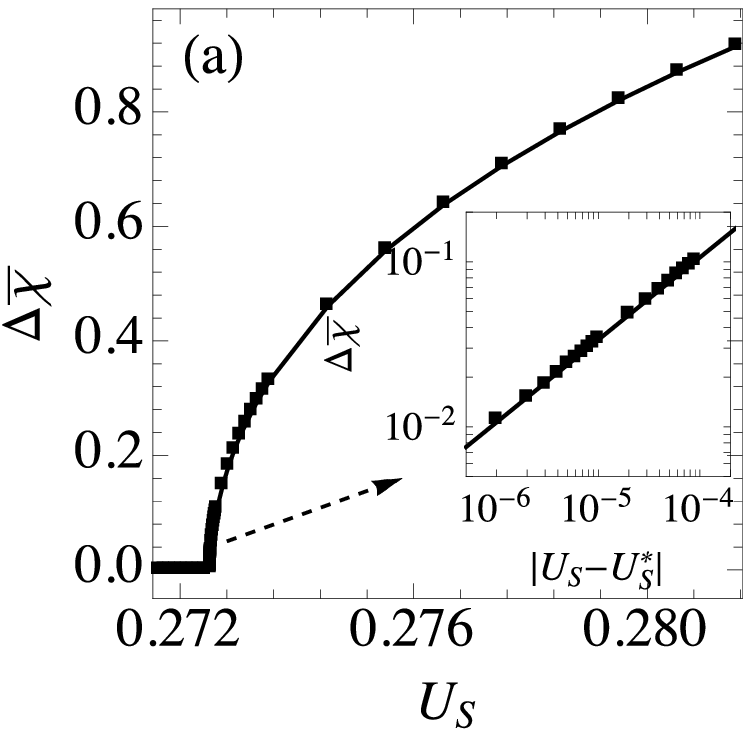}\includegraphics[width=1.6in]{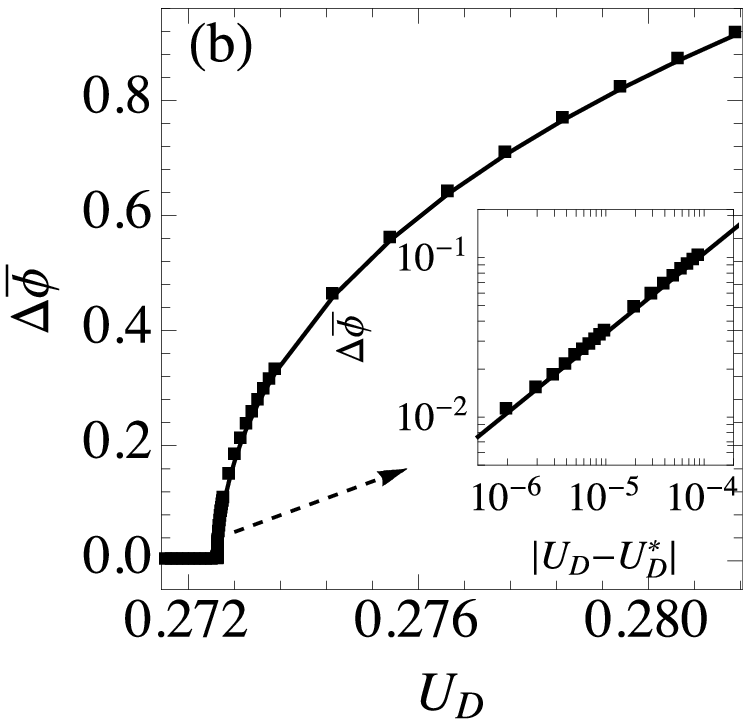}

\includegraphics[width=1.6in]{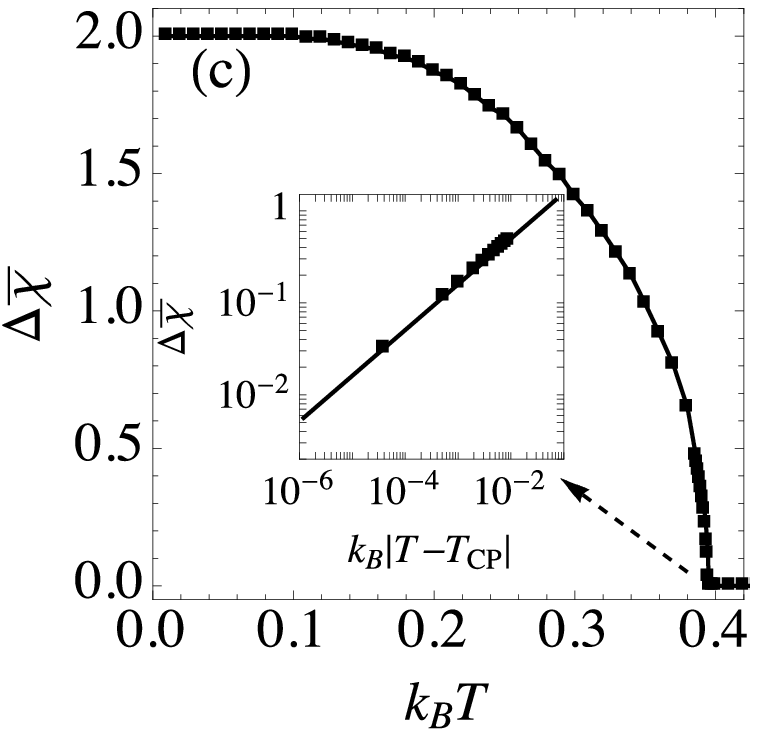}\includegraphics[width=1.6in]{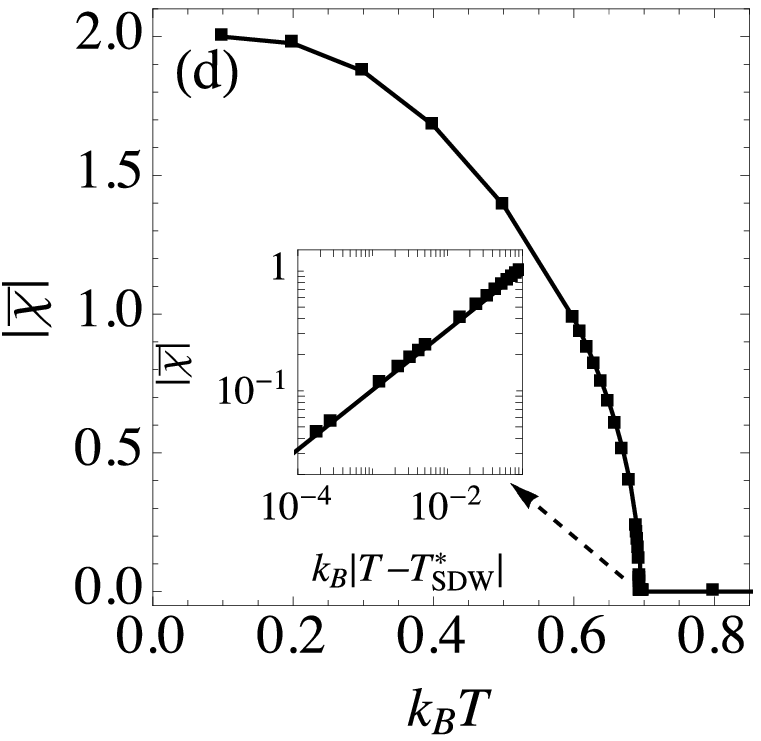}

\caption{Critical scaling behavior of the SDW and CDW order parameter in the
critical regime. (a,~b) Scaling behavior of order parameter jumps
$\Delta\bar{\chi}$ and $\Delta\bar{\phi}$ in the vicinity of the
bicritical point with $k_{B}T=0.5$. The main plots show the interaction
strength dependence of $\Delta\bar{\chi}$ and $\Delta\bar{\phi}$
along the first-order SDW-CDW transition boundary {[}cf.~Fig.~\ref{Fig_1_Phase_diagram}(d){]}.
The insets show the power law fits to the interaction strength dependence
of $\Delta\bar{\phi}$ and $\Delta\bar{\chi}$ in the vicinity of
the bicritical point $(U_{D}^{*},U_{S}^{*})=(0.2725,0.2725)$, clearly
manifesting the power law scaling behavior $\Delta\bar{\chi}\propto|U_{S}-U_{S}^{*}|^{0.499}$
and $\Delta\bar{\phi}\propto|U_{D}-U_{D}^{*}|^{0.499}$. (c) SDW order
parameter jump $\Delta\chi$ versus temperature along the first-order
transition boundary shown in Fig.~\ref{Fig_1_Phase_diagram}(c),
showing a power law scaling $\Delta\bar{\chi}\propto|T-T_{\mathrm{CP}}|^{0.497}$
near the critical point $k_{B}T_{\mathrm{CP}}=0.395$ as shown by
the inset in the plot. (d) Temperature dependence of SDW order parameter
$|\bar{\chi}|$ with $(U_{S},U_{D})$ kept fixed at $(0.3,\,0.2)$
when $T$ is tuned across a second-order homogeneous-SDW transition
point with $k_{B}T_{\mathrm{SDW}}^{*}=0.69525$, showing a power law
scaling of $\propto(T-T_{\mathrm{SDW}}^{*})^{0.496}$ near the second-order
transition as shown by the inset in the plot. See text for more details.}
\label{Fig_2_critial_scaling}
\end{figure}

In this paper, we address this question for two-component lattice
Bose gases in optical cavities. To this end, we establish the finite-temperature
phase diagram of the system in the deep Mott-insulator limit (cf.~Fig.~\ref{Fig_1_Phase_diagram})
and investigate the emergent critical scaling of the system (cf.~Fig.~\ref{Fig_2_critial_scaling}).
More specifically, we find the following. (i) An emergent critical
regime featuring in particular a new bicritical line associated with
the SDW-CDW transition at finite temperature. At low temperatures,
we find transitions between each two of the three phases, namely,
the homogeneous, the SDW, and the CDW phase are all first-order transitions,
where the SDW (CDW) order parameter assumes finite jump $\Delta\bar{\chi}$
($\Delta\bar{\phi}$) when the transition boundary is crossed {[}cf.~Fig.~\ref{Fig_1_Phase_diagram}(b){]}.
When the temperature increases, $\Delta\bar{\chi}$ and $\Delta\bar{\phi}$
for the homogeneous-SDW and the homogeneous-CDW transitions decrease,
finally vanish at their respective critical points, and both transitions
become second-order transitions {[}cf.~Fig.~\ref{Fig_1_Phase_diagram}(c)
and Figs.~\ref{Fig_2_critial_scaling}(c,~d){]}. At the same time,
a new bicritical point emerges, where the first-order SDW-CDW transition
terminates at this point with vanishing order parameter jumps $\Delta\bar{\chi}$
and $\Delta\bar{\phi}$ {[}cf.~Fig.~\ref{Fig_1_Phase_diagram}(d)
and Figs.~\ref{Fig_2_critial_scaling}(a,~b){]}. (ii) The universality
class of the emergent critical scaling belongs to the five-dimensional
(5D) Ising universality class (cf.~Fig.~\ref{Fig_2_critial_scaling}).
This clearly shows that the criticality of the system is strongly
influenced by and thus bear the long-range characteristic of its interactions.

\emph{System and model in the deep Mott-insulator limit.}---Among
current diverse experimental setups of multi-component Bose gases
in optical cavities \citep{Landini_PRL_2018,Morales_Nat_Mat_2018,Kroeze_PRL_2018,Morales_PRA_2019,Dogra_Science_2019,Davis_PRL_2019,Li_arXiv_2020},
here we concentrate on the experimental setups in Ref.~\citep{Landini_PRL_2018,Dogra_Science_2019},
where, in particular, cavity mediated spin-spin interactions were
realized. Moreover, an additional static square optical lattice is
assumed in the system, similar to what has been realized for single-component
Bose gases in optical cavities \citep{Landig_Nature_2016}. For this
type of two-component lattice spinor Bose gases in optical cavities,
their physics in a large parameter regime can be described by an ILR
interacting Hubbard-type Hamiltonian, which consists of a conventional
hopping part and an interaction part (see Supplemental Material \citep{Sup_Mat}
for the derivation of the lattice model). Here, we focus on the physics
in the deep Mott-insulator limit which is completely determined by
the interaction part of the Hamiltonian whose explicit form reads
\begin{align}
\hat{H}= & \frac{1}{2}\sum_{i,\sigma\sigma'}U_{\sigma\sigma'}\hat{n}_{i,\sigma}\left(\hat{n}_{i,\sigma'}-\delta_{\sigma\sigma'}\right)\nonumber \\
 & -\frac{1}{L}\left[U_{D}\left(\hat{N}_{e}-\hat{N}_{o}\right)^{2}+U_{S}\left(\hat{S}_{e}-\hat{S}_{o}\right)^{2}\right].\label{eq:Lattice_model}
\end{align}
The first term of the Hamiltonian (\ref{eq:Lattice_model}) describes
the on-site intra- and inter-component interactions whose strengths
are characterized by $U_{\sigma\sigma'}$ with $\sigma(\sigma')=\pm$
being the component index. The second term in Eq.~(\ref{eq:Lattice_model})
describes the cavity-mediated ILR density-density and spin-spin interactions
with their strengths characterized by $U_{D}$ and $U_{S}$, respectively.
Noticing in this term, both $U_{D}$ and $U_{S}$ are further rescaled
by the total number of lattice sites $L$ according to the Kac prescription
\citep{Kac_J_Math_Phys_1963} in order to restore the conventional
thermodynamical limit. Here, the two interpenetrating square sub-lattices
of the complete square lattice are referred to as ``even'' ($e$)
and ``odd'' ($o$) lattice, respectively. $\hat{n}_{i,\sigma}$
is the particle number operator that counts the number of atoms with
component index $\sigma$ at site $i$. $\hat{N}_{e(o)}\equiv\sum_{i\in e(o)}\left(\hat{n}_{i,+}+\hat{n}_{i,-}\right)$
and $\hat{S}_{e(o)}\equiv\sum_{i\in e(o)}\left(\hat{n}_{i,+}-\hat{n}_{i,-}\right)$
are the total density and total spin operator of the even (odd) sub-lattice. 

From the Hamiltonian (\ref{eq:Lattice_model}) we notice that the
system assumes both a $\mathbb{Z}_{2}$-symmetry with respect to the
two sub-lattices, i.e., exchanging the sub-lattice indices ``$e$''
and ``$o$'', and a $\mathbb{Z}_{2}$-symmetry with respect to the
two-components, i.e., exchanging the component indices ``$-$''
and ``$+$''. The ILR density-density interaction favors a CDW phase
that breaks the $\mathbb{Z}_{2}$-symmetry between the two sub-lattices,
while the ILR spin-spin interaction favors an SDW phase that breaks
both $\mathbb{Z}_{2}$-symmetries. The competition among these two
types of ILR interaction and the short-range on-site interaction gives
rise to phase transitions among the SDW, CDW, and the homogeneous
phase as observed in experiments focusing at fixed low temperatures
\citep{Landini_PRL_2018}. At the temperature scale that is comparable
with the energy scales of interactions, one would expect that thermal
fluctuations could give rise to new physics that is absent in the
low-temperature regime. Indeed, as we shall see in the following,
new criticality associated to phase transitions among the SDW, CDW,
and the homogeneous phase arises in this finite temperature regime. 

\emph{Finite-temperature phase diagram and emergent criticality at
intermediate temperature scale.---}To establish the complete finite-temperature
phase diagram, we directly calculate the quantum grand partition function
$Z=\mathrm{tr}[e^{-\beta(\hat{H}-\mu_{\sigma}\sum_{i}\hat{n}_{i,\sigma})}]$
of the system, which can be formulated as an integral with respect
to the SDW order parameter field $\chi$ and the CDW order parameter
field $\phi$ \citep{Sup_Mat}, and explicitly reads
\begin{align}
Z & =\frac{\beta L}{\pi}\sqrt{U_{D}U_{S}}\iint_{-\infty}^{+\infty}d\chi d\phi\,e^{-\beta L\Omega_{\{U_{S},U_{D},U_{\sigma\sigma'},\mu_{\sigma},\beta\}}(\phi,\chi)},\label{eq:Grand_partition_function}
\end{align}
with 
\begin{align}
 & \Omega_{\{U_{S},U_{D},U_{\sigma\sigma'},\mu_{\sigma},\beta\}}(\phi,\chi)\nonumber \\
\equiv & -\frac{1}{2\beta}\sum_{\eta=\pm1}\ln\left[\sum_{n_{\pm}=0}^{+\infty}e^{-\beta\left(\underset{\sigma\sigma'}{\sum}\frac{1}{2}U_{\sigma\sigma'}n_{\sigma}(n_{\sigma}-\delta_{\sigma\sigma'})-\underset{\sigma}{\sum}\mu_{\sigma}n_{\sigma}\right)}\right.\nonumber \\
 & \left.e^{-2\beta\eta[U_{D}\phi(n_{+}+n_{-})+U_{S}\chi(n_{+}-n_{-})]}\right]+U_{D}\phi^{2}+U_{S}\chi^{2}.\label{eq:Omega_explicit_form}
\end{align}
Here, $\mu_{\sigma}$ is the chemical potential for the species with
component index $\sigma$, and $\beta=(k_{B}T)^{-1}$ with $k_{B}$
being the Boltzmann constant and $T$ being the temperature. Order
parameter fields $\chi$ and $\phi$ in Eq.~(\ref{eq:Grand_partition_function})
are introduced via the standard Hubbard-Stratonovich transformation,
and their expectation values $\bar{\chi}$ and $\bar{\phi}$ correspond
exactly to the CDW and SDW order parameter, respectively, i.e., $\bar{\chi}\equiv\langle\chi\rangle=\langle\hat{S}_{e}-\hat{S}_{o}\rangle/L$,
$\bar{\phi}\equiv\langle\phi\rangle=\langle\hat{N}_{e}-\hat{N}_{o}\rangle/L$.
In the thermodynamic limit $L\rightarrow\infty$, the partition function
$Z$ is exactly determined by its saddle point integration, hence
SDW and CDW order parameters $\bar{\phi}$ and $\bar{\chi}$ are determined
by the value of $(\phi,\chi)$ that minimizes $\Omega_{\{U_{S},U_{D},U_{\sigma\sigma'},\mu_{\sigma},\beta\}}(\phi,\chi)$
(see Supplemental Material \citep{Sup_Mat} for more technical details).
The summation in Eq.~(\ref{eq:Omega_explicit_form}) can be numerically
evaluated at sufficiently high accuracy with a large enough cut-off
on $n_{\sigma}$. This enables us to map out the complete finite-temperature
phase diagram as we shall now discuss.

In the following, we focus on the balanced case with unit filling
for both species of atoms, i.e., $\langle\sum_{i}\hat{n}_{i,\sigma}\rangle/L=1$
for $\sigma=\pm$. In addition, the intra-component interaction strengths
$U_{++}$ and $U_{--}$ are assumed to be equal and much larger than
the inter-component interaction strength $U_{+-}$, i.e., $U_{+-}\ll U_{++}=U_{--}$.
This corresponds to the case where the physics associated with the
internal ``spin'' degrees of freedom is dominated by the ILR spin-spin
interaction. For the convenience of the discussion, the on-site inter-component
interaction $U_{\sigma\sigma}$ is set to be the energy unit in the
following. The finite-temperature phase diagram of the system for
this case is shown in Fig.~\ref{Fig_1_Phase_diagram}(a), which consists
of three phase regions that correspond to the SDW, the CDW, and the
homogeneous phase, respectively. At low temperature, the transitions
between each two of these three phases are all first-order transitions
as shown in Fig.~\ref{Fig_1_Phase_diagram}(b), which is a cross-section
of the phase diagram Fig.~\ref{Fig_1_Phase_diagram}(a) at $k_{B}T=0.1.$
This corroborates recent findings in experiments \citep{Li_arXiv_2020}. 

In the parameter regime where the temperature scale is comparable
to the on-site interaction energy {[}cf.~the vicinal region of the
two red lines and the purple line in Fig.~\ref{Fig_1_Phase_diagram}(a){]},
rich critical behavior appear. This manifests particularly in the
emergence of the two critical lines and the bicritical line {[}marked
by red and purple, respectively, in Fig.~\ref{Fig_1_Phase_diagram}(a){]}. 

The two critical lines consist of critical points at which the first-order
homogenous-SDW transition or the homogeneous-CDW transition terminates
and changes to the second-order transition. For instance, in Fig.~\ref{Fig_1_Phase_diagram}(c),
a cross-section of the phase diagram Fig.~\ref{Fig_1_Phase_diagram}(a)
at $U_{D}=0.2$ is shown, and one can directly observe that the homogeneous-SDW
transition changes from a first-order transition to a second-order
one at the critical point with $T_{\mathrm{CP}}=0.39/k_{B}$ (marked
by the red dot in the plot). In fact, the second-order homogenous-SDW
and homogeneous-CDW transitions above these two critical lines give
rise to a critical line consisting of a new type of critical point
that is absent in the corresponding single-component systems, namely,
the bicritical point {[}cf.~the purple dot in Fig.~\ref{Fig_1_Phase_diagram}(d){]}.

The emergence of the bicritical points can be straightforwardly seen
by monitoring the change of the $U_{S}-U_{D}$ phase diagram of the
system when the temperature is increased, as illustrated by Fig.~\ref{Fig_1_Phase_diagram}(b)
and Fig.~\ref{Fig_1_Phase_diagram}(d) which show two typical $U_{S}-U_{D}$
phase diagrams at relatively low ($k_{B}T=0.1$) and high ($k_{B}T=0.5$)
temperature. We can directly see from these two diagrams that at low
temperatures all the transition boundaries are first-order ones, while
in the temperature regime above the two-critical lines, both the homogeneous-SDW
and homogeneous-CDW transition boundary become the second-order transition
boundary {[}cf.~the two solid lines in Fig.~\ref{Fig_1_Phase_diagram}(d){]}
and change the point where these two boundaries meet to a bicritical
point {[}cf.~the purple dot in Fig.~\ref{Fig_1_Phase_diagram}(d){]},
at which the first-order SDW-CDW transition boundary also terminates.

The emergence of these two critical lines and the bicritical line
gives rise to a new critical regime that is absent at low temperatures
where all the transitions are first-order ones. As we shall see in
the following, in this critical regime, the system manifests critical
power law scalings characteristic of its long-range interactions.

\emph{Critical scaling and universality class of transitions among
SDW, CDW, and homogeneous phases at finite temperatures.---}For the
SDW-CDW transition, irrespective of the temperature, it is always
a first-order transition, i.e., both the SDW and CDW order parameter
show finite jump when either $U_{S}$ or $U_{D}$ is tuned across
the first-order SDW-CDW transition boundary {[}cf.~Fig.~\ref{Fig_1_Phase_diagram}(b,~d){]}.
Consequentially, neither $\phi$ nor $\chi$ shows any critical power
law scaling. However, in the temperature regime where the bicritical
point emerges, the SDW-CDW transition can indeed assume a type of
critical scaling in the vicinity of the bicritical point, manifesting
in both order parameter jumps, namely, $\Delta\bar{\chi}$ and $\Delta\bar{\phi}$.
As we can see from Figs.~\ref{Fig_2_critial_scaling}(a,~b), which
show the ILR interaction strength dependence of $\Delta\bar{\chi}$
and $\Delta\bar{\phi}$ along a first-order SDW-CDW transition boundary
that terminates at a bicritical point {[}cf.~Fig.~\ref{Fig_1_Phase_diagram}(d){]},
both $\Delta\bar{\chi}$ and $\Delta\bar{\phi}$ decrease monotonously
with respect to the IRL interaction strengths and finally vanish at
the bicritical point. Power law fits to the interaction strength dependence
of $\Delta\bar{\chi}$ and $\Delta\bar{\phi}$ in the vicinity of
the bicritical point $(U_{D}^{*},U_{S}^{*})=(0.2725,0.2725)$ {[}cf.~the
purple dot in Fig.~\ref{Fig_1_Phase_diagram}(d){]} as shown in the
insets of Figs.~\ref{Fig_2_critial_scaling}(a,~b), clearly show
the power law critical scaling behavior $\Delta\bar{\chi}\propto|U_{S}-U_{S}^{*}|^{0.499}$
and $\Delta\bar{\phi}\propto|U_{D}-U_{D}^{*}|^{0.499}$.

For the homogeneous-SDW and the homogeneous-CDW transition, there
are two types of critical scaling behavior associated with each of
them. The first type manifests in the order parameter jump in the
vicinity of the critical point where the first-order transition change
to a second-order one {[}cf.~Fig.~\ref{Fig_1_Phase_diagram}(c){]}.
Let us take the homogeneous-SDW transition for instance. As we can
see from Fig.~\ref{Fig_2_critial_scaling}(c), which shows the temperature
dependence of $\Delta\bar{\chi}$ along a first-order homogeneous-SDW
transition boundary that terminates at a critical point {[}cf.~Fig.~\ref{Fig_1_Phase_diagram}(c){]},
$\Delta\chi$ decrease monotonously with respect to temperature and
finally vanish at the critical point. Power law fit to the temperature
dependence of $\Delta\bar{\chi}$ in the vicinity of the critical
point $k_{B}T_{\mathrm{CP}}=0.395$ {[}cf.~the red dot in Fig.~\ref{Fig_1_Phase_diagram}(c){]}
as shown in the inset of Fig.~\ref{Fig_2_critial_scaling}(c) clearly
shows the power law critical scaling behavior $\Delta\bar{\chi}\propto|T-T_{\mathrm{CP}}|^{0.497}$.
The second type of critical scaling behavior manifests in the order
parameter in the vicinity of the second-order transition. Let us still
take the homogeneous-SDW transition for instance. As we can see from
Fig.~\ref{Fig_2_critial_scaling}(d), which shows the temperature
dependence of $|\bar{\chi}|$ when $(U_{S},U_{D})$ is kept fixed
at $(0.3,\,0.2)$, and $T$ is tuned across a second-order homogeneous-SDW
transition point with $k_{B}T_{\mathrm{SDW}}^{*}=0.69525$ {[}cf.~Fig.~\ref{Fig_1_Phase_diagram}(c){]},
$|\bar{\chi}|$ decrease monotonously with respect to the temperature
and finally vanish at the transition point. Power law fit to the temperature
dependence of $|\bar{\chi}|$ in the vicinity of the second-order
transition point as shown in the inset of Fig.~\ref{Fig_2_critial_scaling}(d)
clearly shows the power law critical scaling behavior $|\bar{\chi}|\propto|T-T_{\mathrm{SDW}}^{*}|^{0.496}$. 

Interestingly, one can notice that the power law scaling behavior
for different transitions manifest almost the same critical exponent
with the value $1/2$. This strongly suggests they should originate
from the same effective critical theory. Indeed, as we shall see below,
all these scaling behavior can be well-described by an effective Ginzburg-Landau
(GL) theory with a double $\mathbb{Z}_{2}$ symmetry.

\begin{figure}
\includegraphics[width=1.7in]{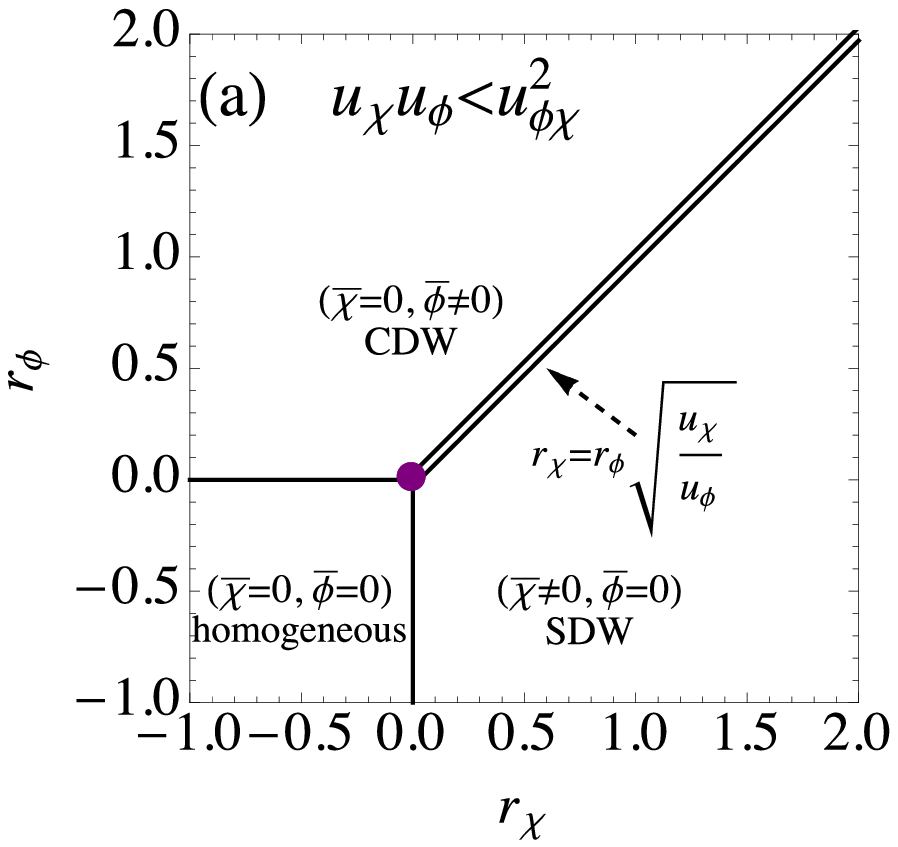}\includegraphics[width=1.7in]{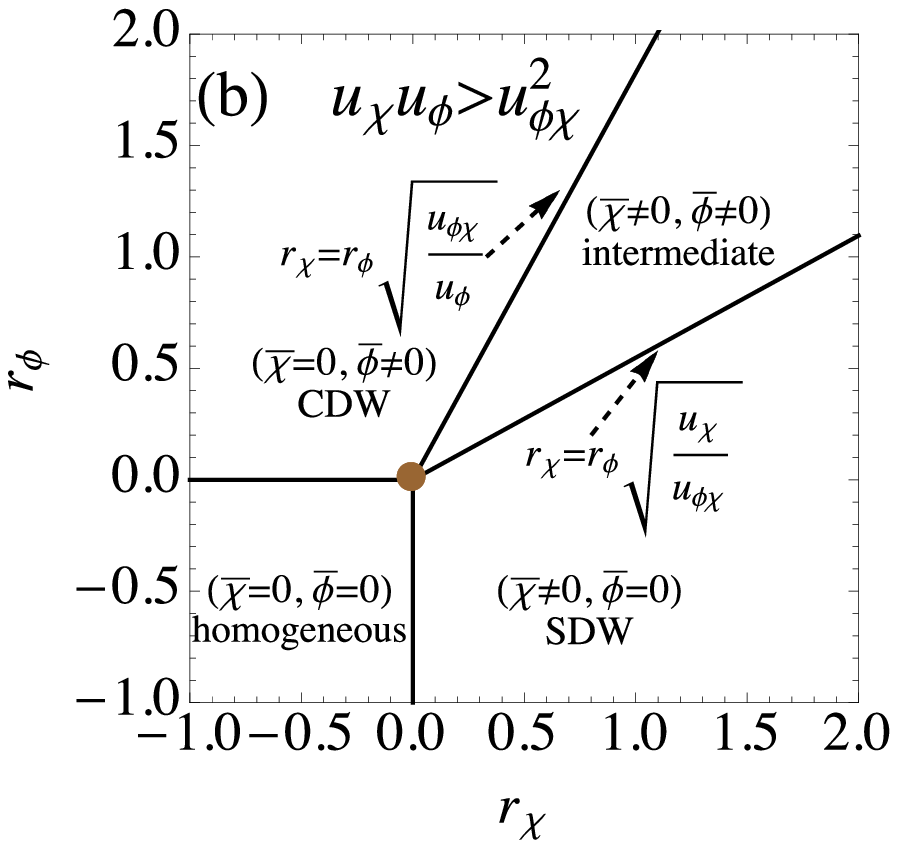}

\caption{Phase diagram of the effective GL theory (\ref{eq:GL_effective_theory})
with double $\mathbb{Z}_{2}$ symmetry that shows either a bicritical
point or a tetracritical point, depending on the sign of $u_{\phi\chi}^{2}-u_{\phi}u_{\chi}$.
Solid curves stand for the second-order transition boundaries, while
the double solid ones stand for the first-order transition boundaries.
Left panel: For $u_{\phi\chi}^{2}>u_{\phi}u_{\chi}$, the phase diagram
support a bicritical point (marked by the purple dot) at $(r_{\phi}=0,r_{\chi}=0)$
where two second-order phase transition boundaries meet. Right panel:
For $u_{\phi\chi}^{2}<u_{\phi}u_{\chi}$, the phase diagram support
a tetracritical point (marked by the brown dot) at $(r_{\phi}=0,r_{\chi}=0)$
where four second-order phase transition boundaries meet. See text
for more details.}
\label{Fig_3_bicritical_point_and_tetracritical_point_GL_theory}
\end{figure}

In the critical regime, both the CDW and the SDW order parameters
are small enough to allow a systematic expansion of the system's free
energy $F$ with respect to them. The double $\mathbb{Z}_{2}$ symmetry
of the system determine the allowed terms in the expansion, whose
explicit form reads

\begin{equation}
F=-\frac{1}{2}r_{\phi}\bar{\phi}^{2}-\frac{1}{2}r_{\chi}\bar{\chi}^{2}+\frac{1}{4}u_{\phi}\bar{\phi}^{4}+\frac{1}{4}u_{\chi}\bar{\chi}^{4}+\frac{1}{2}u_{\phi\chi}\bar{\phi}^{2}\bar{\chi}^{2},\label{eq:GL_effective_theory}
\end{equation}
with $r_{\phi}$, $r_{\chi}$, $u_{\phi}$, $u_{\chi}$, and $u_{\phi\chi}$
being the GL coefficients. Comparing to the single-component case,
the novel aspect of physics, in this case, is the competition between
the two ordered phases, namely, the CDW and the SDW phase, that is
driven by the relative strength between $U_{D}$ and $U_{S}$. In
the spirit of Ginzburg-Landau effective theory, to describe this scenario,
we assume that $r_{\phi}\propto(U_{D}-U_{D}^{*})$ and $r_{\chi}\propto(U_{S}-U_{S}^{*})$
with $(U_{D}^{*},U_{S}^{*})$ being the critical point around which
both order parameters, i.e., $\bar{\phi}$ and $\bar{\chi}$, are
small. $u_{\phi},\,u_{\chi},$ and $u_{\phi\chi}$ are positive fixed
GL parameters that do not depend on the tuning parameter $U_{D}$
and $U_{S}$. By analyzing the saddle points of $F$ (see Supplemental
Material \citep{Sup_Mat} for analysis details), depending on the
sign of $u_{\phi\chi}^{2}-u_{\phi}u_{\chi}$, one can find two distinct
phase diagrams for the effective theory (\ref{eq:GL_effective_theory})
as shown in Fig.~\ref{Fig_3_bicritical_point_and_tetracritical_point_GL_theory}.
Comparing to results from direct calculations as shown in Fig.~\ref{Fig_1_Phase_diagram}(d),
one naturally expects the system under consideration is described
by the GL theory with $u_{\phi}u_{\chi}<u_{\phi\chi}^{2}$, where
there is a direct first-order transition between the SDW and CDW phases.
Direct calculations within the GL theory with $u_{\phi}u_{\chi}<u_{\phi\chi}^{2}$
show that the order parameter jumps $\Delta\bar{\phi}$ and $\Delta\bar{\chi}$
along the first-order SDW-CDW transition line assumes the forms $\Delta\bar{\phi}=\sqrt{r_{\phi}/u_{\phi}}$
and $\Delta\bar{\chi}=\sqrt{r_{\chi}/u_{\chi}}$ \citep{Sup_Mat}.
Noticing $r_{\phi}$ and $r_{\chi}$ depend linearly on interaction
strengths, one thus directly obtains the following scaling law 
\begin{equation}
\Delta\bar{\chi}\propto|U_{S}-U_{S}^{*}|^{1/2},\,\Delta\bar{\phi}\propto|U_{D}-U_{D}^{*}|^{1/2}.\label{eq:Scaling_bicritical_point}
\end{equation}
We remark here that from the phase diagrams of the effective theory
(\ref{eq:GL_effective_theory}), the tetracritical point is in principle
allowed by the GL theory with the double $\mathbb{Z}_{2}$ symmetry.
Interestingly, in the experimental setup with two-component Bose gases
in the presence of two optical resonators \citep{Morales_Nat_Mat_2018},
this tetracritical point is indeed observed. This indicates, for the
type of experimental systems in Ref.~\citep{Morales_Nat_Mat_2018},
its GL effective theory is the one with $u_{\phi}u_{\chi}>u_{\phi\chi}^{2}$.
To discuss the scaling behavior associated with the homogeneous-SDW
transition or the homogeneous-CDW transition, since either CDW or
the SDW order is identically zero in the transition under consideration,
the effective theory (\ref{eq:GL_effective_theory}) in fact degenerates
to the one with the single $\mathbb{Z}_{2}$ symmetry and can be straightforwardly
analyzed as what has been done in the single-component case \citep{He_arXiv_2020_1}.
Take the homogeneous-SDW transition, for instance, this analysis results
in power law scaling $\Delta\bar{\chi}\propto|T-T_{\mathrm{CP}}|^{1/2}$
in the vicinity of the critical point, and $|\bar{\chi}|\propto|T-T_{\mathrm{SDW}}^{*}|^{1/2}$
in the vicinity of the second-order transition. In fact, from the
structure of the phase diagram Fig.~\ref{Fig_3_bicritical_point_and_tetracritical_point_GL_theory}(a),
it is straightforward to see that the scaling behavior for $\Delta\bar{\chi}$
($\Delta\bar{\phi}$) with respect to the interaction strength $U_{S}$
($U_{D}$) in Eq.~(\ref{eq:Scaling_bicritical_point}) is the same
as the one for $\bar{\chi}$ ($\bar{\phi}$) of the second-order homogeneous-SDW
(homogeneous-CDW) transition, which is determined by the effective
GL theories with the single $\mathbb{Z}_{2}$ symmetry. Therefore,
all the scaling exponents discussed above is governed essentially
by the same effective GL theories with the single $\mathbb{Z}_{2}$
symmetry.

By comparing the critical exponents from the GL effective theory and
the ones from direct calculations, one immediately notices remarkable
agreement that seems counter-intuitive at first sight, since long-range
fluctuations are omitted in the effective GL theory, and it is only
expected to provide rough estimations of the critical exponents for
the 2D system under consideration. In fact, this good agreements between
the mean-field type effective GL theory and direct exact calculations,
originate from the fact that long-range fluctuations in the critical
regime are strongly suppressed by the ILR interactions in the system,
hence making effective GL theory a precise theory in the critical
regime. Such a similar promotion of an effective GL theory to a precise
effective critical theory is also identified in the single-component
Bose gases in optical cavities \citep{He_arXiv_2020_1}. Noticing
the corresponding critical exponent of the 5D Ising model is exactly
$1/2$ \citep{Aizenman_PRL_1981} and can be obtained by the same
effective GL theory, this thus concludes that the emergent criticality
of the system at finite temperature belongs to the 5D Ising universality
class, manifests clearly the long-range character of its interactions.

\emph{Conclusions.---}Thermal fluctuations at intermediate temperature
regime can strongly influence the competition between magnetic and
density order of multi-component Bose gases in optical cavities, giving
rise to rich critical behavior: The first-order SDW-homogeneous and
CDW-homogeneous transition become second-order transitions, and at
the same time giving rise to a new bicritical line, where the first-order
SDW-CDW transition terminates at this line with vanishing order parameter
jumps. The critical scaling behavior in this critical regime belong
to the five-dimensional Ising universality class, clearly characterizing
the long-range nature of the system's interactions. With current well-established
experimental techniques for detecting the SDW and CDW order \citep{Landig_Nature_2016,Landini_PRL_2018},
we expect our findings can be directly observed by current experimental
setups \citep{Landini_PRL_2018} operated at the temperature scale
around half of the on-site energy. We believe our work will stimulate
further experimental and also theoretical investigations on possible
emergent critical behavior in multicomponent Bose gases in optical
cavities in the presence of thermal fluctuations, particularly beyond
the deep Mott-insulator limit.
\begin{acknowledgments}
This work was supported by NSFC (Grant No.~11874017, No.~11674334,
and No.~11947302), GDSTC under Grant No.~2018A030313853, Science
and Technology Program of Guangzhou (Grant No.~2019050001), and START
grant of South China Normal University.
\end{acknowledgments}

\end{document}

\renewcommand{\theequation}{S-\arabic{equation}} \renewcommand{\thefigure}{S\arabic{figure}} \renewcommand{\thetable}{S\arabic{table}}
\title{Supplementary Material for ``Emergent criticality and universality
class of spin and charge density wave transitions of two-component
lattice Bose gases in optical cavities at finite temperature'' }
\author{Liang He}
\email{liang.he@scnu.edu.cn}

\affiliation{Guangdong Provincial Key Laboratory of Quantum Engineering and Quantum
Materials, SPTE, South China Normal University, Guangzhou 510006,
China}
\author{Su Yi}
\email{syi@itp.ac.cn}

\affiliation{CAS Key Laboratory of Theoretical Physics, Institute of Theoretical
Physics, Chinese Academy of Sciences, Beijing 100190, China}
\affiliation{School of Physical Sciences \& CAS Center for Excellence in Topological
Quantum Computation, University of Chinese Academy of Sciences, Beijing
100049, China}
\maketitle

\section*{Hamiltonian of two-component lattice bose gases in optical cavities}

In current experiments, multicomponent Bose gases are loaded in the
free space inside cavities. We consider here a closely related system
where a two-component Bose gas is loaded in a two dimensional (2D)
square optical lattice within an optical cavity. As we shall present
in the following, the Hamiltonian of this system can be derived from
the atom-photon interacting model that corresponds to the experimental
setup in Ref.~\citep{Landini_PRL_2018}.

Our starting point is the single-particle atom-photon interacting
model for two-component or pseudo-spin $1/2$ bosonic atoms in a 2D
square optical lattice which reads
\begin{eqnarray}
\hat{H}_{\mathrm{sp}} & = & \left\{ \frac{\left|\mathbf{\hat{p}}\right|^{2}}{2m}+V_{2\mathrm{D}}[\cos^{2}(k\hat{x})+\cos^{2}(k\hat{z})]\right\} \otimes\mathbb{I}\nonumber \\
 &  & +\lambda_{D}(\hat{a}^{\dagger}+\hat{a})\cos(k\hat{x})\cos(k\hat{z})\otimes\mathbb{I}\nonumber \\
 &  & -i\lambda_{S}(\hat{a}^{\dagger}-\hat{a})\cos(k\hat{x})\cos(k\hat{z})\otimes\sigma_{z}\nonumber \\
 &  & -\Delta_{c}\hat{a}^{\dagger}\hat{a}+U_{0}\hat{a}^{\dagger}\hat{a}\cos^{2}(k\hat{x})\otimes\mathbb{I}\label{eq:single_particle_atom-photon_model}
\end{eqnarray}
where $\lambda_{D}$ ($\lambda_{S}$) is the spin-independent (spin-dependent)
atom-photon interaction strength, $V_{2D}$ is the strength of the
2D optical lattice potential, $\hat{a}^{\dagger}$ ($\hat{a}$) is
the creation (annihilation) operator for the photons in the cavity,
$\Delta_{c}$ is the detuning between the frequency of the cavity
photon and the frequency of the $z-$lattice beam, $U_{0}$ is the
maximum light shift per atom, $\mathbb{I}$ is the $2\times2$ identity
matrix, and $\sigma_{z}$ is the conventional Pauli matrix. After
taking the lowest band approximation, we can directly write down the
corresponding many-body atom-photon Hamiltonian in the complete second
quantization form, namely
\begin{eqnarray}
 &  & \hat{H}_{\mathrm{AP}}\\
 & = & \int d\mathbf{r}\hat{\Psi}^{\dagger}(\mathbf{r})\left[-\frac{\hbar\nabla^{2}}{2m}+V_{2\mathrm{D}}(\cos^{2}(kx)+\cos^{2}(kz))\right]\hat{\Psi}(\mathbf{r})\nonumber \\
 & + & \int d\mathbf{r}\hat{\Psi}^{\dagger}(\mathbf{r})\left[\lambda_{D}(\hat{a}^{\dagger}+\hat{a})\cos(kx)\cos(kz)\right]\hat{\Psi}(\mathbf{r})\nonumber \\
 & + & \int d\mathbf{r}\hat{\Psi}^{\dagger}(\mathbf{r})\left[-i\lambda_{S}(\hat{a}^{\dagger}-\hat{a})\cos(kx)\cos(kz)\right]\sigma_{z}\hat{\Psi}(\mathbf{r})\nonumber \\
 & + & \left(\int d\mathbf{r}\hat{\Psi}^{\dagger}(\mathbf{r})U_{0}\cos^{2}(kx)\hat{\Psi}(\mathbf{r})-\Delta_{c}\right)\hat{a}^{\dagger}\hat{a}\nonumber 
\end{eqnarray}
where $\hat{\Psi}(\mathbf{r})\equiv\left(\sum_{\mathbf{i}}W_{\mathbf{i}}(x,z)\hat{b}_{\mathbf{i,+}},\,\sum_{\mathbf{i}}W_{\mathbf{i}}(x,z)\hat{b}_{\mathbf{i,-}}\right)^{T}$
with $W_{\mathbf{i}}(x,z)$ being the Wannier function on the site
$\mathbf{i}$ in the lowest band and $\hat{b}_{\mathbf{i,\pm}}$ being
its corresponding annihilation operator for the atom with component
index $\sigma=\pm$. After further taking into account the contact
interaction between atoms, the system can be well described by the
following many-body Hamiltonian 
\begin{eqnarray}
 &  & \hat{H}_{\mathrm{AP}}\label{eq:H_AP}\\
 & = & \frac{1}{2}\sum_{\mathbf{i},\sigma\sigma'}U_{\sigma\sigma'}\hat{n}_{\mathbf{i},\sigma}\left(\hat{n}_{\mathbf{i},\sigma'}-\delta_{\sigma\sigma'}\right)-t\sum_{\langle\mathbf{i},\mathbf{j}\rangle,\sigma}\left(\hat{b}_{\mathbf{i},\sigma}^{\dagger}\hat{b}_{\mathbf{j},\sigma}+\mathrm{H}.\mathrm{c}.\right)\nonumber \\
 &  & +\lambda_{D}(\hat{a}^{\dagger}+\hat{a})\left(\hat{N}_{e}-\hat{N}_{o}\right)-i\lambda_{S}(\hat{a}^{\dagger}-\hat{a})\left(\hat{S}_{e}-\hat{S}_{o}\right)\nonumber \\
 &  & -\left(\Delta_{c}-\delta\right)\hat{a}^{\dagger}\hat{a}\nonumber 
\end{eqnarray}
where $U_{\sigma\sigma'}$ is the contact interaction strength between
atoms with component index $\sigma$ and $\sigma'$, $t$ is the hopping
amplitude, and $\delta$ is a dispersive shift \citep{Landig_Nature_2016}.
Here, $\hat{N}_{e(o)}=\sum_{\mathbf{i}\in e(o)}\left(\hat{n}_{\mathbf{i},+}+\hat{n}_{\mathbf{i},-}\right)$
and $\hat{S}_{e(o)}=\sum_{\mathbf{i}\in e(o)}\left(\hat{n}_{\mathbf{i},+}-\hat{n}_{\mathbf{i},-}\right)$,
with $\hat{n}_{\mathbf{i},\sigma}\equiv\hat{b}_{\mathbf{i},\sigma}^{\dagger}\hat{b}_{\mathbf{i},\sigma}$. 

To derive the effective Hamiltonian for the atoms only, we first write
down the Heisenberg equation of motion for cavity photons with a finite
decay rate $\kappa$, i.e., $id\hat{a}/dt=[\hat{a},\hat{H}_{\mathrm{AP}}]-i\kappa\hat{a}$,
whose explicit form reads
\begin{eqnarray}
i\frac{d\hat{a}}{dt} & = & -\left(\Delta_{c}-\delta\right)\hat{a}-i\kappa\hat{a}\\
 &  & +\lambda_{D}\left(\hat{N}_{e}-\hat{N}_{o}\right)-i\lambda_{S}\left(\hat{S}_{e}-\hat{S}_{o}\right).\nonumber 
\end{eqnarray}
Typically, $\kappa$ is much larger than the atomic recoil energy
scale, thus we can approximate $\hat{a}$ as 
\begin{equation}
\hat{a}=\frac{\lambda_{D}\left(\hat{N}_{e}-\hat{N}_{o}\right)-i\lambda_{S}\left(\hat{S}_{e}-\hat{S}_{o}\right)}{\Delta_{c}-\delta+i\kappa}.
\end{equation}
Plugging the above expression into Eq.~(\ref{eq:H_AP}), we thus
eliminate the cavity field adiabatically and obtain the effective
Hubbard-type Hamiltonian $\hat{H}_{\mathrm{Hub}}$ for the atoms only,
i.e., 
\begin{align}
\hat{H}_{\mathrm{Hub}} & =-t\sum_{\langle\mathbf{i},\mathbf{j}\rangle,\sigma}\left(\hat{b}_{\mathbf{i},\sigma}^{\dagger}\hat{b}_{\mathbf{j},\sigma}+\mathrm{H}.\mathrm{c}.\right)\\
 & +\frac{1}{2}\sum_{\mathbf{i},\sigma\sigma'}U_{\sigma\sigma'}\hat{n}_{\mathbf{i},\sigma}\left(\hat{n}_{\mathbf{i},\sigma'}-\delta_{\sigma\sigma'}\right)\nonumber \\
 & -\frac{1}{L}\left[U_{l,D}\left(\hat{N}_{e}-\hat{N}_{o}\right)^{2}+U_{l,S}\left(\hat{S}_{e}-\hat{S}_{o}\right)^{2}\right],\nonumber 
\end{align}
with 
\begin{align}
U_{l,D} & \equiv-2L\lambda_{D}^{2}\frac{\Delta_{c}-\delta}{(\Delta_{c}-\delta)^{2}+\kappa^{2}},\\
U_{l,S} & \equiv-2L\lambda_{S}^{2}\frac{\Delta_{c}-\delta}{(\Delta_{c}-\delta)^{2}+\kappa^{2}}.
\end{align}

\section*{Partition function after the hubbard-strtonovich transformation }

The grand quantum partition function for the system in the Mott-insulator
limit reads
\begin{align}
Z & =\mathrm{tr}\left[e^{-\beta(\hat{H}-\mu_{\sigma}\hat{N}_{\sigma})}\right]\\
 & =\sum_{\{n_{i,\sigma}\}}e^{-\beta\left(\sum_{i,\sigma,\sigma'}\frac{1}{2}U_{\sigma\sigma'}(n_{i,\sigma}-\delta_{\sigma\sigma'})\right)}\nonumber \\
 & \times e^{-\beta\left(-L^{-1}\left[U_{D}\left(N_{e}-N_{o}\right)^{2}+U_{S}\left(S_{e}-S_{o}\right)^{2}\right]\right)},\nonumber 
\end{align}
where $N_{e(o)}=\sum_{i\in e(o)}\left(n_{i,+}+n_{i,-}\right)$ and
$S_{e(o)}=\sum_{i\in e(o)}\left(n_{i,+}-n_{i,-}\right)$, with $n_{i,\sigma}$
being the eigenvalue of the particle number operator $\hat{n}_{i,\sigma}$.
To treat the long-range density-density and spin-spin interaction
term, we introduce two Hubbard-Stratonovich transformations in the
density and spin channel, respectively, i.e.,
\begin{align}
 & \left(\sqrt{\frac{\beta U_{D}L}{\pi}}\right)^{-1}\exp\left(\beta\frac{U_{D}}{L}\left[N_{e}-N_{o}\right]^{2}\right)\label{eq:HST_Den}\\
 & =\int_{-\infty}^{+\infty}d\phi\,\exp\left(-\beta U_{D}\left\{ L\phi^{2}+2\left[N_{e}-N_{o}\right]\phi\right\} \right),\nonumber \\
 & \left(\sqrt{\frac{\beta U_{S}L}{\pi}}\right)^{-1}\exp\left(\beta\frac{U_{S}}{L}\left[S_{e}-S_{o}\right]^{2}\right)\label{eq:HST_Spi}\\
 & =\int_{-\infty}^{+\infty}d\chi\,\exp\left(-\beta U_{S}\left\{ L\chi^{2}+2\left[S_{e}-S_{o}\right]\chi\right\} \right).\nonumber 
\end{align}
Plugging the above transformations into the partition function, we
arrive at 
\begin{align}
Z & =\frac{\beta L}{\pi}\sqrt{U_{D}U_{S}}\iint_{-\infty}^{+\infty}d\chi d\phi\,e^{-\beta L\Omega_{\{U_{S},U_{D},U_{\sigma\sigma'},\mu_{\sigma},\beta\}}(\phi,\chi)},\label{eq:Grand_partition_function}
\end{align}
where 
\begin{align}
 & \Omega_{\{U_{S},U_{D},U_{\sigma\sigma'},\mu_{\sigma},\beta\}}(\phi,\chi)\\
\equiv & -\frac{1}{2\beta}\sum_{\eta=\pm1}\ln\left[\sum_{n_{\pm}=0}^{+\infty}e^{-\beta\left(\underset{\sigma\sigma'}{\sum}\frac{1}{2}U_{\sigma\sigma'}n_{\sigma}(n_{\sigma}-\delta_{\sigma\sigma'})-\underset{\sigma}{\sum}\mu_{\sigma}n_{\sigma}\right)}\right.\nonumber \\
 & \left.e^{-2\beta\eta[U_{D}\phi(n_{+}+n_{-})+U_{S}\chi(n_{+}-n_{-})]}\right]+U_{D}\phi^{2}+U_{S}\chi^{2}.\nonumber 
\end{align}
Here $\chi$ and $\phi$ assume the physical meaning of the fluctuating
SDW and CDW order parameter field, respectively. Their expectation
values $\bar{\chi}\equiv\langle\chi\rangle$ and $\bar{\phi}\equiv\langle\phi\rangle$
equal to the SDW and CDW order parameter, respectively, i.e., $\bar{\chi}=\langle\hat{S}_{e}-\hat{S}_{o}\rangle/L$,
$\bar{\phi}=\langle\hat{N}_{e}-\hat{N}_{o}\rangle/L$. 

\section*{Analysis of the Ginzburg-Landau effective theory with a double-$\mathbb{Z}_{2}$
symmetry}

Since both the CDW order parameter $\bar{\phi}$ and the SDW order
parameter $\bar{\chi}$ are small in the critical regime, one can
perform a systematic expansion of the system's free energy with respect
to these two order parameters. Up to the fourth order in the order
parameters, the double-$\mathbb{Z}_{2}$ symmetry allowed form of
the Ginzburg-Landau (GL) free energy $F$ reads 
\begin{equation}
F=-\frac{1}{2}r_{\phi}\bar{\phi}^{2}-\frac{1}{2}r_{\chi}\bar{\chi}^{2}+\frac{1}{4}u_{\phi}\bar{\phi}^{4}+\frac{1}{4}u_{\chi}\bar{\chi}^{4}+\frac{1}{2}u_{\phi\chi}\bar{\phi}^{2}\bar{\chi}^{2}.\label{eq:GL_effective_theory}
\end{equation}

Now let us discuss the interaction-induced transition between the
CDW and SDW phase at a fixed temperature that is above the critical
temperature which separates the first-order homogeneous-SDW (homogeneous-CDW)
transition from the second-order homogeneous-SDW (homogeneous-CDW)
transition. In the spirit of Ginzburg-Landau effective theory, we
assume that $r_{\phi}\propto(U_{D}-U_{D}^{*})$ and $r_{\chi}\propto(U_{S}-U_{S}^{*})$
with $(U_{D}^{*},U_{S}^{*})$ being the critical point around which
both order parameters, i.e., $\bar{\phi}$ and $\bar{\chi}$, are
small. $u_{\phi},\,u_{\chi},$ and $u_{\phi\chi}$ are positive fixed
GL parameters that do not depend on the tuning parameter $U_{D}$
and $U_{S}$. The saddle point of GL free energy is determined by
the equations $\partial F/\partial\bar{\phi}=0$ and $\partial F/\partial\bar{\chi}=0$,
whose explicit form reads 
\begin{align}
\bar{\phi}\left(-r_{\phi}+u_{\phi}\bar{\phi}^{2}+u_{\phi\chi}\bar{\chi}^{2}\right) & =0,\\
\bar{\chi}\left(-r_{\chi}+u_{\chi}\bar{\chi}^{2}+u_{\phi\chi}\bar{\phi}^{2}\right) & =0.
\end{align}
By straightforwardly investigating the solution of the above saddle
point equations, one can directly find the phase digram of the GL
effective theory (\ref{eq:GL_effective_theory}) can assume two distinct
structures depending on the sign of $u_{\phi\chi}^{2}-u_{\phi}u_{\chi}$. 

In the case with $u_{\phi\chi}^{2}-u_{\phi}u_{\chi}>0$, the phase
diagram assumes two second-order phase transition boundaries that
correspond to the homogeneous-SDW and homogeneous-CDW transition and
one first-order transition boundary that corresponds to the SDW-CDW
transition. The phase with both orders does not exist. These two second-order
phase transition boundaries meet at a bicritical point $(r_{\phi}=0,r_{\chi}=0)$
at which the order parameter jumps associated with the first-order
SDW-CDW transition vanish, i.e., the first-order SDW-CDW transition
boundary terminates at this point. The order parameter jumps assume
the forms $\Delta\bar{\phi}=\sqrt{\frac{r_{\phi}}{u_{\phi}}}$ and
$\Delta\bar{\chi}=\sqrt{\frac{r_{\chi}}{u_{\chi}}}$, from which one
can conclude the corresponding power law scaling $\Delta\bar{\phi}\propto|U_{D}-U_{D}^{*}|^{1/2},\Delta\bar{\chi}\propto|U_{S}-U_{S}^{*}|^{1/2}$. 

While for the case with $u_{\phi\chi}^{2}-u_{\phi}u_{\chi}<0$, there
exists an intermediate phase that assumes both the non-zero SDW and
the non-zero CDW order parameter, whose explicit forms read 
\begin{equation}
\chi^{2}=\frac{u_{\phi\chi}r_{\phi}-r_{\chi}u_{\phi}}{u_{\phi\chi}^{2}-u_{\phi}u_{\chi}},\,\phi^{2}=\frac{u_{\phi\chi}r_{\chi}-r_{\phi}u_{\chi}}{u_{\phi\chi}^{2}-u_{\phi}u_{\chi}},
\end{equation}
from which one can directly show that the phase boundary of this intermediate
phase regime is determined by 
\begin{equation}
r_{\phi}\frac{u_{\phi\chi}}{u_{\phi}}\leq r_{\chi}\leq r_{\phi}\frac{u_{\chi}}{u_{\phi\chi}}.
\end{equation}
The transition from this intermediate phase to either the SDW or the
CDW phase is the second-order transition. Moreover, both the homogeneous-SDW
and homogeneous-CDW transitions are second-order transitions. From
the phase diagram of the GL effective theory, we see that all these
four different second-order transition lines meet at the same point
$(r_{\phi}=0,r_{\chi}=0)$, giving rise to the tetracritical point.